\documentstyle[12pt]{article}

\begin{document}

\renewcommand{\theequation}{\thesection.\arabic{equation}}
\newcommand{\reseteqnum}{\setcounter{equation}{0}}

\title{
\hfill
\parbox{4cm}{\normalsize UT-KOMABA 98-26\\hep-th/9811258}\\
\vspace{2cm}
D-branes on Three-dimensional\\
Nonabelian Orbifolds
\vspace{1.5cm}}
\author{Tomomi Muto\thanks{e-mail address:
\tt muto@hep1.c.u-tokyo.ac.jp}
\vspace{0.5cm}\\
{\normalsize\em Institute of Physics}\\
{\normalsize\em University of Tokyo, Komaba, Tokyo 153}}
\date{\normalsize}
\maketitle
\vspace{1cm}

\begin{abstract}
\normalsize

We study D-branes on a three complex dimensional
nonabelian orbifold ${\bf C}^3/\Gamma$ with $\Gamma$
a finite subgroup of $SU(3)$.
We present general formulae necessary to obtain
quiver diagrams which represent
the gauge group and the spectrum of the D-brane
worldvolume theory for dihedral-like subgroups
$\Delta(3n^2)$ and $\Delta(6n^2)$.
It is found that the quiver diagrams have a similar
structure to webs of branes.
\end{abstract}

\newpage
\section{Introduction}
D-branes serve as probes to study short distance structure
of spacetime. From a D-brane point of view, spacetime
emerges from moduli space of D-brane worldvolume gauge
theory and spacetime
coordinates are promoted to noncommuting matrices.
It is very different from geometric pictures
based on general relativity and fundamental strings,
so it is natural to ask whether various aspects
of spacetime are modified or not if we probe spacetime by D-branes.
Investigations in this direction have been developed in recent
years.
Especially, D-branes on orbifolds have been intensively studied.

D-branes on an orbifold ${\bf C}^2/\Gamma$ with $\Gamma$ a finite
subgrop of $SU(2)$ were investigated in \cite{DM,JM}.
In this case, D-brane worldvolume gauge theory has ${\cal N}=2$
supersymmetry.
Finite subgroups of $SU(2)$ are classified into $ADE$ series:
$A$-type subgroups are abelian, while $D$-type and $E$-type
subgroups are nonabelian.
The gauge group and the spectrum of the D-brane
gauge theory are represented by a quiver diagram,
which corresponds to the Dynkin diagram
of $ADE$ affine Lie algebra.

For the ${\bf C}^3/\Gamma$ case with $\Gamma$ a finite
subgrop of $SU(3)$, D-brane worldvolume gauge theory
has ${\cal N}=1$ supersymmetry.
Abelian orbifolds ($\Gamma={\bf Z}_n,{\bf Z}_n
\times {\bf Z}_{n'}$) were investigated in
\cite{DGM,Muto,Douglas,MR}.
The phase structure of the Kahler muduli space is investigated
and it is shown that only geometric phases appear by using toric
methods.
It is also shown that topology changing process
occurs as in the analyses based on fundamental strings
\cite{AGM,Witten}.
On the other hand, nonabelian cases have not been studied so far.
But it is known that finite subgroups of $SU(3)$ have a similar
classification to the $ADE$ classification in the $SU(2)$ case
\cite{FFK}.
That is, finite subgroups of $SU(3)$ other than $SU(2)$ and direct
products of abelian phase groups fall into 2 series:
\begin{itemize}
\item
The analogues of dihedral subgroups of $SU(2)$ ($D$-type); \\
$\Delta(3n^2)$ \, ($n$ is a positive integer) \\
$\Delta(6n^2)$ \, ($n$ is a positive even integer)
\item
The analogues of exceptional subgroups of $SU(2)$ ($E$-type); \\
$\Sigma(60)$,\, $\Sigma(168)$,\, $\Sigma(360x)$,\,
$\Sigma(36x)$,\,$\Sigma(72x)$,\,$\Sigma(216x)$ \, $(x=1,3)$
\end{itemize}
Here the number in braces is the order of the group.

The purpose of the present paper is to investigate D-branes on
${\bf C}^3/\Gamma$ with $\Gamma$ a nonabelian finite subgroup of
$SU(3)$, and study what kinds of gauge groups and matter contents
are allowed for the D-brane worldvolume gauge thory.
However, when we were preparing this manuscript,
we received papers \cite{HH,GLR} which treat similar problems.
The complete list of gauge groups and matter contents of
D-brane world volume theory
for $\Sigma$-type subgroups is given in \cite{HH},
so, in this paper, we concentrate on $\Delta$-type subgroups.

The organization of this paper is as follows.
In section 2, we review a prescription to obtain
worldvolume gauge theory of a D-brane on ${\bf C}^3/\Gamma$.
In section 3, we present general formulae which are necessary
to obtain the spectrum of the gauge theory of the D-brane on
${\bf C}^3/\Gamma$ with $\Gamma=\Delta(3n^2)$ and $\Delta(6n^2)$.
It is also pointed out that the formulae have a structure that resembles
the condition for a three string junction\cite{Schwarz}
or a web of $(p,q)$ 5-branes\cite{AHK}.
Section 4 contains discussions.
Group theoretical properties of $\Delta(3n^2)$ and $\Delta(6n^2)$
are given in the appendix.

\section{Orbifold projection}
\reseteqnum

In this section, we recapitulate the prescription
to construct D-brane worldvolume gauge theory
on an orbifold ${\bf C}^3/\Gamma$.
We start with $N$ parallel D3-branes on ${\bf C}^3$ where
$N=|\Gamma|$ is the order of $\Gamma$.
The D-brane worldvolume theory is ${\cal N}=4$ supersymmetric
4-dimensional $U(N)$ gauge theory.
The bosonic field content are three complex adjoint scalars
$X^\mu$ ($\mu=1,2,3$) and a $U(N)$ gauge field $A$.
Then we project this
theory into $\Gamma$ invariant space.
This condition is expressed as
\begin{eqnarray}
R_{reg} A R_{reg}^{-1} &=& A, \label{eq:projectionA}\\
(R_3)_{\mu\nu} R_{reg} X^\nu R_{reg}^{-1} &=& X^\mu
\label{eq:projectionX}
\end{eqnarray}
where $R_{reg}$ is the $N \times N$ regular representation
which acts on Chan-Paton index
and $R_3$ is a 3-dimensional representation
which acts on spacetime index $\mu$.
$R_3$ defines how $\Gamma$ acts on ${\bf C}^3$
to form the quotient singularity.
The regular representation $R_{reg}$ has the following form
\begin{equation}
R_{reg}=\oplus_{a=1}^r N_a R^a
\end{equation}
where $R^a$ is an irreducible representation,
$N_a={\rm dim}R^a$ and $r$ is the number of irreducible
representations.
Due to the condition (\ref{eq:projectionA}),
gaauge symmetry of the projected theory becomes
\begin{equation}
\prod_{a=1}^r U(N_a).
\end{equation}
The chiral matter obtained after the projection (\ref{eq:projectionX})
can be found by computing the tensor product of the 3-dimensional
representation $R_3$ and an irreducible representation $R^a$,
\begin{equation}
R_3 \otimes R^a = \oplus_{b=1}^r n_{ab}^{\bf 3} R^b.
\label{eq:tensor}
\end{equation}
$n_{ab}^{\bf 3}$ represents the number of  fields which
transforms as
$N_a \otimes \bar N_b$ under $U(N_a) \times U(N_b)$.

The gauge group and the spectrum
are summarized in a quiver diagram.
A quiver diagram consists of $r$ nodes
and arrows which connect these nodes:
$n_{ab}^{\bf 3}$ represents the number of arrows
from the $a$-th node to the $b$-th node.
So once we  calculate the coefficient $n_{ab}^{\bf 3}$,
we can obtain the field content.

Now we briefly review how to calculate the coeficient
$n_{ab}^{\bf 3}$.
A reducible representation $R_{red}$ is decomposed
into a direct sum of irreducible representations,
\begin{equation}
R_{red} = \oplus_{a=1}^r n_a R^a.
\label{eq:reduction}
\end{equation}
We denote the character for an element $g \in \Gamma$
in a representation $R^i$ as $\chi^i(g)$,
then we have the following equation
corresponding to the decomposition (\ref{eq:reduction}),
\begin{equation}
\chi^{red}(g) = \sum_{a=1}^r n_a \chi^a(g).
\end{equation}
By using the orthogonality condition for the
irreducible representations
\begin{equation}
\frac{1}{|\Gamma|} \sum_{g \in \Gamma}
\chi^a(g) \chi^b(g)^* = \delta_{ab},
\end{equation}
we can express the coeficient $n_a$ in the equation
(\ref{eq:reduction}) as
\begin{equation}
n_a = \frac{1}{|\Gamma|} \sum_{g \in \Gamma} \chi(g) \chi^a(g)^*.
\label{eq:multiplicity}
\end{equation}
For a direct product of representations
\begin{equation}
R^i \otimes R^j = R^k,
\end{equation}
the following relation holds
\begin{equation}
\chi^i(g) \chi^j(g) = \chi^k(g).
\label{eq:product}
\end{equation}
Combining equations (\ref{eq:multiplicity}) and
(\ref{eq:product}), the coefficient $n_{ab}^{\bf 3}$
in the tensor product (\ref{eq:tensor}) is expressed as
\begin{equation}
n_{ab}^{\bf 3} = \frac{1}{|\Gamma|} \sum_{g \in \Gamma}
\chi^3(g) \chi^a(g) \chi^b(g)^*.
\label{eq:multiplicity2}
\end{equation}
The elements of $\Gamma$ are classified into conjugacy classes.
Characters are common for all elements in the same
conjugacy class, so the expression (\ref{eq:multiplicity2})
can be rewritten as
\begin{equation}
n_{ab}^{\bf 3} = \frac{1}{|\Gamma|} \sum_{c=1}^r
|C_c| \chi^3(C_c) \chi^a(C_c) \chi^b(C_c)^*.
\end{equation}
where $r$ is the number of conjugacy classes,
which is the same as the number of irreducible representations,
and $|C_a|$ is the number of elements of the class $C_a$.

\section{D-branes on ${\bf C}^3/\Delta(3n^2)$ and
${\bf C}^3/\Delta(6n^2)$}
\reseteqnum

In this section, we present formulae for tensor products
which represent the spectrum of the D-brane gauge theory
on ${\bf C}^3/\Delta(3n^2)$ and ${\bf C}^3/\Delta(6n^2)$.
We discuss physical implications of the formulae.
Group theoretical properties of these subgrops are summarized
in the appendix.

\subsection{$\Delta (3n^2)$ case}

We first consider the case with $n \notin 3{\bf Z}$.
In this case, the irreducible representations consists of
three 1-dimensional representations
and $(n^2-1)/3$ 3-dimensional representations.
So the gauge group of the corresponding worldvolume theory is
$U(1)^3 \times U(3)^{(n^2-1)/3}$.
We denote the 1-dimensional representations as
$R_1^\alpha$ $(\alpha=0,1,2)$,
where $R_1^0$ is the trivial representation.
The 3-dimensional representations are labeled by
two integers $(m_1,m_2)$ with $(m_1,m_2) \neq (0,0)$.
These integers are defined modulo $n$ and furthermore
there are equivalence relations among the representations
$R_3^{(m_1,m_2)}$,
\begin{equation}
R_3^{(m_1,m_2)}=R_3^{(-m_1+m_2,-m_1)}
=R_3^{(-m_2,m_1-m_2)}.
\label{eq:relation}
\end{equation}
So we can restrict the region of $(m_1,m_2)$ as follows.
\begin{equation}
0 \leq 2 m_2-m_1 < n,\quad
-n < m_2-2m_1 \leq 0
\label{eq:region}
\end{equation}
and $(m_1,m_2) \neq (0,0)$.
Hence $R_3^{(0,0)}$ is not an irreducible 3-dimensional
representation.
Instead, we define
\begin{equation}
R_3^{(0,0)} \equiv R_1^1 \oplus
R_1^2 \oplus R_1^3.
\label{eq:definition}
\end{equation}
Then tensor products of $R_3^{(m_1,m_2)}$ and
the irreducible representations are given by a rather compact form,
\begin{eqnarray}
&&R^{(m_1,m_2)}_3 \otimes R^{\alpha}_1
= R^{(m_1,m_2)}_3, \label{eq:a1} \\
&&R^{(m_1,m_2)}_3 \otimes R^{(l_1,l_2)}_3
= R^{(l_1+m_1,l_2+m_2)}_3
\oplus R^{(l_1-m_2,l_2+m_1-m_2)}_3
\oplus R^{(l_1-m_1+m_2,l_2-m_1)}_3. \label{eq:a2}
\end{eqnarray}
As noted in the last section,
$R^{(l_1,l_2)}_3$ is taken to be an irreducible representation,
hence, $(l_1,l_2) \neq (0,0)$.
However we can take $l_i$ to be any integer
since the second equation (\ref{eq:a2}) with
$(l_1,l_2)=(0,0)$ is consisent with the first one (\ref{eq:a1})
under the definition (\ref{eq:definition}).
Therefore the structure of the tensor products and the
corresponding quiver diagram is essentially expressed by
the second equation (\ref{eq:a2}).

As we noted earlier, the quiver diagram consists of nodes
associated with the irreducible representations
and arrows associated with the matter contents.
In the present case, the irreducible representations
are labeled by two integers modulo $n$,
so we can put the nodes on a lattice ${\bf Z}_n \times {\bf Z}_n$
with some equivalence relations.
The equation (\ref{eq:a2}) shows that there are three arrows
which start from $(l_1,l_2)$.
The terminal points are $(l_1+m_1,l_2+m_2)$,
$(l_1-m_1+m_2,l_2-m_1)$ and
$(l_1-m_2,l_2-m_1+m_2)$.
This part of the quiver diagram is given in figure 1.
The total quiver diagram is obtained by putting such arrows
to each nodes in ${\bf Z}_n \times {\bf Z}_n$
and identifying the nodes according to the equivalence relations
(\ref{eq:relation}).
(To be exact, the node at $(0,0)$ must be splitted into three nodes
due to the equation (\ref{eq:definition}).)

\begin{figure}[htb]
\begin{center}
\begin{picture}(200,120)
%\put(0,0){\framebox(200,120)}
\put(165,103){$(l_1+m_1,l_2+m_2)$}
\put(-65,90){$(l_1-m_2,l_2+m_1-m_2)$}
\put(85,10){$(l_1-m_1+m_2,l_2-m_1)$}
\put(105,60){$(l_1,l_2)$}
\put(100,70){\circle*{3}}
\thicklines
\put(100,70){\vector(3,2){60}}
\put(100,70){\vector(-2,1){40}}
\put(100,70){\vector(-1,-3){20}}
\end{picture}
\caption{Parts of the quiver diagram for ${\bf C}^3/\Delta(3n^2)$.}
\end{center}
\end{figure}
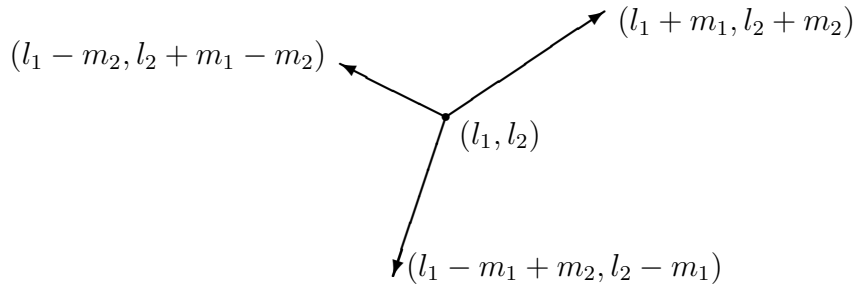

Now we comment on the structure of the quiver diagram.
The figure 1 graphically resembles a three string junction
\cite{Schwarz} or a web of $(p,q)$ 5-branes\cite{AHK}.
In fact, relations between junctions of $(p,q)$ strings
and representations of algebras were clarified in \cite{DZ}
for the cases of $ADE$ Lie algebras.
In \cite{LV}, it was also pointed out that
local geometry of a certain Calabi-Yau threefold
has a realization in terms of a web of $(p,q)$ 5-branes.
So it is quite interesting to make the connection
between quiver diagrams, brane configurations and
geometric singularities more definite
for the cases considered in this paper.

Here it is worth mentioning to a relation to so-called
brane box models\cite{HU}.
The finite subgroup $\Delta(3n^2)$ (and $\Delta(6n^2)$)
can be understood as a generalization of ${\bf Z}_n \times
{\bf Z}_n$.
So the orbifolds studied in this paper are generalizations
of orbifolds ${\bf C}^3/{\bf Z}_n \times {\bf Z}_n$.
For a D-brane on ${\bf C}^3/{\bf Z}_n \times {\bf Z}_n$,
we have a counterpart of the equation (\ref{eq:a2}),
\[
R^{(m_1,m_2)}_3 \otimes R^{(l_1,l_2)}_1
= R^{(l_1+m_1,l_2+m_2)}_1
\oplus R^{(l_1-m_1,l_2)}_1
\oplus R^{(l_1,l_2-m_2)}_1. \label{eq:a3}
\]
The corresponding part of the quiver diagram is given in
figure 2.

\begin{figure}[htp]
\begin{center}
\begin{picture}(200,120)
%\put(0,0){\framebox(200,120)}
\put(165,93){$(l_1+m_1,l_2+m_2)$}
\put(105,20){$(l_1,l_2-m_2)$}
\put(-25,60){$(l_1-m_1,l_2)$}
\put(105,52){$(l_1,l_2)$}
\put(100,60){\circle*{3}}
\thicklines
\put(100,60){\vector(3,2){60}}
\put(100,60){\vector(-1,0){60}}
\put(100,60){\vector(0,-1){40}}
\end{picture}
\caption{Parts of the quiver diagram for ${\bf C}^3/{\bf Z}_n
\times {\bf Z}_n$.}
\end{center}
\end{figure}
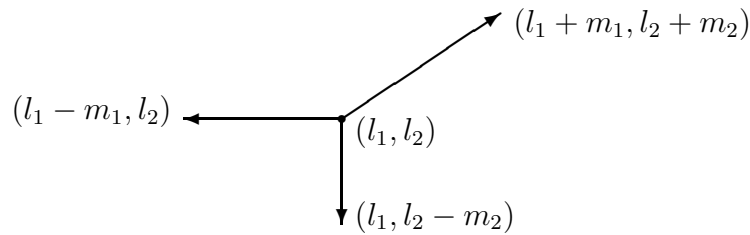

The brane box is a model that gives the same gauge theory
as the D-brane worldvolume theory on ${\bf C}^3/{\bf Z}_n
\times {\bf Z}_n$.
In the brane box model,
the figure 2 is directly related to a brane configuration;
two types of NS-branes cross orthogonally to each other.
So it is natural to expect that the figure 1 represents
some brane configuration in which branes cross each
other at a certain angle.
It would be interesting to study brane configurations from this viewpoint.

Next we consider the case where $n \in 3{\bf Z}$.
In this case, the irreducible representations consists of
nine 1-dimensional representations
and $n^2/3-1$ 3-dimensional representations.
So the gauge group is
$U(1)^9 \times U(3)^{n^2/3-1}$.
We denote the 1-dimensional representations as
$R_1^\alpha$ $(\alpha=0,1,\dots, 8)$,
where $R_1^0$ is the trivial representation.
The 3-dimensional representations are labeled by
two integers $(m_1,m_2)$ where $(m_1,m_2) \neq (0,0),
(n/3,2n/3),(2n/3,n/3)$.
As in the $n \notin 3{\bf Z}$ case, there are equivalence
relations (\ref{eq:relation}),
so we can restrict the region of $(m_1,m_2)$ as (\ref{eq:region}).
By defining
\begin{eqnarray}
&&R_3^{(0,0)} \equiv R_1^0 \oplus
R_1^1 \oplus R_1^2,\\
&&R_3^{(n/3,2n/3)} \equiv R_1^3 \oplus
R_1^4 \oplus R_1^5,\\
&&R_3^{(2n/3,n/3)} \equiv R_1^6 \oplus
R_1^7 \oplus R_1^8,
\end{eqnarray}
tensor products of $R_3^{(m_1,m_2)}$ and
the irreducible representations are given as follows,
\begin{eqnarray}
&&R^{(m_1,m_2)}_3 \otimes R^{0,1,2}_1
= R^{(m_1,m_2)}_3,\\
&&R^{(m_1,m_2)}_3 \otimes R^{3,4,5}_1
= R^{(n/3+m_1,2n/3+m_2)}_3,\\
&&R^{(m_1,m_2)}_3 \otimes R^{6,7,8}_1
= R^{(2n/3+m_1,n/3+m_2)}_3,\\
&&R^{(m_1,m_2)}_3 \otimes R^{(l_1,l_2)}_3
= R^{(l_1+m_1,l_2+m_2)}_3
\oplus R^{(l_1-m_2,l_2+m_1-m_2)}_3
\oplus R^{(l_1-m_1+m_2,l_2-m_1)}_3.\label{eq:b4}
\end{eqnarray}
Again the first three equations are consistent
with the last one, so the structure of the tensor product
is essentially expressed by the last equation (\ref{eq:b4}).
Although the expression is the same as the equation
(\ref{eq:a2}) in the $n \notin 3 {\bf Z}$ case,
the structure is different.
For example, if we take $m_1+m_2 \in 3 {\bf Z}$,
all the arrows in the quiver diagram have components
of the form $(m,-m+3{\bf Z})$ with $m \in {\bf Z}$.
In this case, not all the nodes are connected
in contrast to the $n \notin 3{\bf Z}$ case.

\subsection{$\Delta (6n^2)$ case}
We now turn to the subgroup $\Delta (6n^2)$.
We first consider the case where $n \notin 3{\bf Z}$.
In this case, the irreducible representations consists of
two 1-dimensional representations,
one 2-dimensional representation,
$2(n-1)$ 3-dimensional representations
and
$(n^2-3n+2)/6$ 6-dimensional representations.
So the gauge group is
$U(1)^2 \times U(2) \times U(3)^{2(n-1)}
\times U(6)^{(n^2-3n+2)/6}$.
We denote the 1-dimensional representations as
$R_1^\alpha$ $(\alpha=0,1)$,
the 2-dimensional representation as $R_2$
and the 3-dimensional representations as
$R_3^{(m,m)t}$, where $m=1,2,\dots, n-1$
and $t$ takes values in ${\bf Z}_2$.
The 6-dimensional representations are labeled by
two integers $(m_1,m_2)$ where $(m_1,m_2) \neq (n,n),
(n,0),(0,n)$.
As  in the $\Delta (3n^2)$ case,
these integers are defined modulo $n$ and there are relations
among the representations $R^{(m_1,m_2)}_6$,
\begin{eqnarray}
&&R_6^{(m_1,m_2)}=R_6^{(-m_1+m_2,-m_1)}
=R_6^{(-m_2,m_1-m_2)} \nonumber\\
&&=R_6^{(m_2,m_1)}=R_6^{(-m_2+m_1,-m_2)}
=R_6^{(-m_1,m_2-m_1)}.
\label{eq:relation6}
\end{eqnarray}
So we can restrict the region of $(m_1,m_2)$ as follows.
\begin{equation}
2m_2 - m_1 \geq 0, \quad m_2 - 2m_1 > -n, \quad
m_1 - m_2 > 0.
\label{eq:region6}
\end{equation}
By defining
\begin{eqnarray}
&&R_3^{(0,0)t} \equiv \delta_{t0} R_1^0 \oplus
\delta_{t1} R_1^1 \oplus R_2,\\
&&R_6^{(m,m)}
\equiv R_3^{(m,m)0} \oplus R_3^{(m,m)1},
\end{eqnarray}
tensor products of $R_3$ and
the irreducible representations are given by a rather compact form,
\begin{eqnarray}
&&R^{(m,m)t}_3 \otimes R^\alpha_1
= R^{(m,m)t+\alpha}_3,\\
&&R^{(m,m)t}_3 \otimes R_2
=R^{(m,m)}_6,\\
&&R^{(m,m)t}_3 \otimes R^{(l,l)t'}_3
= R^{(l+m,l+m)t+t'}_3 \oplus
R^{(l,l-m)}_6,\\
&&R^{(m,m)t}_3 \otimes R^{(l_1,l_2)}_6
= R^{(l_1+m,l_2+m)}_6 \oplus R^{(l_1,l_2-m)}_6
\oplus R^{(l_1-m,l_2)}_6, \label{eq:c4}
\end{eqnarray}
As in the $\Delta(3n^2)$ case,
the structure of the tensor products (and the associated
quiver diagram) is essentially expressed by the last
equation (\ref{eq:c4}).
This equation has a similar structure to the $\Delta(3n^2)$ case
except the equivalence relations among the representations.

Next we consider the case where $n \in 3{\bf Z}$.
In this case, the irreducible representations consists of
two 1-dimensional representations,
four 2-dimensional representations,
$2(n-1)$ 3-dimensional representations
and $(n^2-3n)/6$ 6-dimensional representations.
So the gauge group is
$U(1)^2 \times U(2)^4 \times U(3)^{2(n-1)}
\times U(6)^{(n^2-3n)/6}$.
We denote the 2-dimensional representations as $R_2^\alpha$
$(\alpha=0,\dots,3)$
and use the same notations as the $n \notin 3 {\bf Z}$ case
for other representations.
As in the $n \notin 3{\bf Z}$ case, there are equivalence
relations (\ref{eq:relation6}),
so we can restrict the region of $(m_1,m_2)$ as (\ref{eq:region6}).
By defining
\begin{eqnarray}
&&R_3^{(0,0)t} \equiv \delta_{t0} R_1^0 \oplus
\delta_{t1} R_1^1 \oplus R_2^0,\\
&&R_6^{(m,m)}
\equiv R_3^{(m,m)0} \oplus R_3^{(m,m)1},\\
&&R_6^{(2n/3,n/3)}
\equiv R_2^1 \oplus R_2^2 \oplus R_2^3,
\end{eqnarray}
tensor products of $R_3$ and
the irreducible representations are given as follows,
\begin{eqnarray}
&&R^{(m,m)t}_3 \otimes R^\alpha_1
= R^{(m,m)t+\alpha}_3,\\
&&R^{(m,m)t}_3 \otimes R_2^0
=R^{(m,m)}_6,\\
&&R^{(m,m)t}_3 \otimes R_2^{1,2,3}
=R_6^{(2n/3+m,n/3+m)},\\
&&R^{(m,m)t}_3 \otimes R^{(l,l)t'}_3
= R^{(l+m,l+m)t+t'}_3 \oplus
R^{(l,l-m)}_6,\\
&&R^{(m,m)t}_3 \otimes R^{(l_1,l_2)}_6
= R^{(l_1+m,l_2+m)}_6 \oplus R^{(l_1,l_2-m)}_6
\oplus R^{(l_1-m,l_2)}_6, \label{eq:d5}
\end{eqnarray}
Again, the structure of the tensor product
is essentially expressed by the last equation (\ref{eq:d5}).

\section{Discussion}
\reseteqnum

In this paper, we have considered a D3-brane on an
nonabelian orbifold
${\bf C}^3/\Gamma$ with $\Gamma = \Delta(3n^2)$ and
$\Delta(6n^2)$, finite subgroups of $SU(3)$,
which leads to 4-dimentional ${\cal N}=1$ supersymmetric
gauge theory. There are many interesting applications.

First, if we take $\Gamma \subset SU(4)$ instead of $SU(3)$,
we obtain non-supersymmetric gauge theories.
In \cite{KS,LNV}, it is argued that such theories lead to
conformal field theories motivated by AdS/CFT correspondence.
Much still remains to be done for the non-supersymmetric case.

Secondly, as we discussed in section 3,
the models considered in this paper are related to the
brane box models or webs of $(p,q)$ 5-branes.
Various correspondences between geometric information and
brane configurations have been discussed.
For example, topology changing process in the geometric picture
corresponds to an exchange of branes in the brane configuration
picture\cite{Uranga}.
On the other hand,
topology change is discussed in \cite{GLR} based on D-branes on
nonabelian orbifolds.
It is shown that phase structure of Kahler moduli
space of nonabelian orbifolds can be partly studied by using
toric methods.
So it is interesting to investigate whether the D-brane
gauge theories considered in this paper are realized
by using brane configuration or not, and if possible,
compare the topology changing process between the orbifold picture
and the brane configuration picture.

Finally, we mention to a relation to Mckay
crrespondence\cite{Mckay}.
In the case of $\Gamma \subset SU(2)$,
quiver diagrams coincide with Dynkin diagrams of $ADE$ affine Lie
algebra, which has a relation to WZW models for $\widehat {SU(2)}$.
In \cite{HH}, it is argued that a similar relation holds for the
$\Gamma \subset SU(3)$ case.
We hope that the formulae given in this paper are helpful to
investigations along this line.

\vskip 1cm
\centerline{\large\bf Acknowledgements}

This work is supported in part by Japan Society for the
Promotion of Science(No. 10-3815).

\vskip 2cm

\appendix
\noindent
{\Large\bf Appendix}
\reseteqnum
\vskip 0.5cm

In this appendix, we tabulate characters for
dihedral-like finite subgroups of $SU(3)$,
$\Delta(3n^2)$ and $\Delta(6n^2)$.

\section{$\Delta(3n^2)$}

The group $\Delta(3n^2)$ with $n$ a positive integer
is given by the following elements,
\[
A_{i,j}=\left(
\begin{array}{ccc}
\omega_n^i&0&0 \\
0&\omega_n^j&0 \\
0&0&\omega_n^{-i-j}
\end{array}
\right)\quad
C_{i,j}=\left(
\begin{array}{ccc}
0&0&\omega_n^i \\
\omega_n^j&0&0 \\
0&\omega_n^{-i-j}&0
\end{array}
\right)\quad
E_{i,j}=\left(
\begin{array}{ccc}
0&\omega_n^i&0 \\
0&0&\omega_n^j \\
\omega_n^{-i-j}&0&0
\end{array}
\right)
\label{eq:ACE}
\]
where $\omega_n=e^{2\pi i/n}$ and $0 \leq i,j <n$.
In the following expressions, if $i$ (or $j$)
is out of this region,
it must be replaced by $i+n {\bf Z}$ so that
$0 \leq i+n {\bf Z} < n$.

\vskip 0.5cm

\begin{center}
\begin{tabular}{|c|c|c|c|c|}
\multicolumn{5}{c}
{Character table for
$\Delta(3n^2) \quad (n \notin 3{\bf Z})$}\\
\hline
&$C_1$&$C_2^{ij}$&$C_3$&$C_4$ \\ \hline
$|C_a|$&1&3&$n^2$&$n^2$ \\ \hline
$\#$classes&1&$(n^2-1)/3$&1&1 \\ \hline
$R_1^k$&1&1&$\omega_3^k$&$\omega_3^{2k}$ \\ \hline
$R_3^{(m_1,m_2)}$&3&$\omega_n^{m_1 i +m_2 j}
+\omega_n^{-m_1 (i+j)+m_2 i}+\omega_n^{m_1 j -m_2 (i+j)}$&0&0 \\
\hline
\end{tabular}
$(0 \leq k \leq 2,\,
0 \leq 2 m_2-m_1 < n,\,
-n < m_2-2m_1 \leq 0,\,
(m_1,m_2) \neq (0,0))$
\begin{eqnarray*}
&&C_1 = \{A_{0,0}\} \\
&&C_2^{ij} = \{A_{i,j}, A_{-i-j,i}, A_{j,-i-j}\}
\quad (n<2i+j<2n, \, n<i+2j<2n) \\
&&C_3 = \{C_{i,j}:0 \leq i,j < n\} \\
&&C_4 = \{E_{i,j}:0 \leq i,j < n\}
\end{eqnarray*}
\end{center}

\vskip 0.5cm

\begin{center}
\begin{tabular}{|c|c|c|c|c|}
\multicolumn{5}{c}
{Character table for
$\Delta(3n^2) \quad (n \in 3{\bf Z})$}\\
\hline
&$C_1^l$&$C_2^{ij}$&$C_3^l$&$C_4^l$ \\ \hline
$|C_a|$&1&3&$n^2/3$&$n^2/3$ \\ \hline
$\#$classes&3&$n^2/3-1$&3&3 \\ \hline
$R_1^k$&1&1&$\omega_3^k$&$\omega_3^{2k}$ \\ \hline
$R_1^{k+3}$&1&$\omega_3^{i-j}$&$\omega_3^{k+l}$&
$\omega_3^{2k+l}$ \\ \hline
$R_1^{k+6}$&1&$\omega_3^{j-i}$&$\omega_3^{k+2l}$&
$\omega_3^{2k+2l}$ \\ \hline
$R_3^{(m_1,m_2)}$&$3\omega_3^{l(m_1+m_2)}$&
$\omega_n^{m_1 i +m_2 j}+\omega_n^{-m_1 (i+j)+m_2 i}
+\omega_n^{m_1 j -m_2 (i+j)}$&0&0 \\ \hline
\end{tabular}
$(0 \leq k \leq 2,\,
0 \leq 2 m_2-m_1 < n,\,
-n < m_2-2m_1 \leq 0,\,
(m_1,m_2) \neq (0,0))$
\begin{eqnarray*}
&&C_1^l = \{A_{ln/3,ln/3}\} \quad (l=0,1,2) \\
&&C_2^{ij} = \{A_{i,j}, A_{-i-j,i}, A_{j,-i-j}\}
\quad (n<2i+j<2n, \, n<i+2j<2n) \\
&&C_3^l = \{C_{i+3{\bf Z}+l,i}:0\leq i < n\} \quad (l=0,1,2) \\
&&C_4^l = \{E_{i+3{\bf Z}+l,i}:0\leq i < n\} \quad (l=0,1,2)
\end{eqnarray*}
\end{center}

\section{$\Delta(6n^2)$}

The group $\Delta(6n^2)$ with $n$ a positive even integer
is given by $A_{i,j}$, $C_{i,j}$, $E_{i,j}$
and the following matrices.
\[
B_{i,j}=\left(
\begin{array}{ccc}
\omega_n^i&0&0 \\
0&0&\omega_n^j \\
0&\omega_n^{n/2-i-j}&0
\end{array}
\right)\,
D_{i,j}=\left(
\begin{array}{ccc}
0&\omega_n^i&0 \\
\omega_n^j&0&0 \\
0&0&\omega_n^{n/2-i-j}
\end{array}
\right)\,
F_{i,j}=\left(
\begin{array}{ccc}
0&0&\omega_n^i \\
0&\omega_n^j&0 \\
\omega_n^{n/2-i-j}&0&0
\end{array}
\right)\nonumber
\label{eq:BDF}
\]

\vskip 0.5cm

\begin{center}
\begin{tabular}{|c|c|c|c|c|c|}
\multicolumn{6}{c}
{Character table for
$\Delta(6n^2) \quad (n \notin 3{\bf Z})$}\\
\hline
&$C_1$&$C_2^i$&$C_3^{ij}$&$C_4$&$C_5^l$ \\ \hline
$|C_a|$&1&3&6&$2n^2$&$3n$ \\ \hline
$\#$classes&1&$n-1$&$(n^2-3n+2)/6$&1&$n$ \\ \hline
$R_1^t$&1&1&1&1&$(-1)^t$ \\ \hline
$R_2$&2&2&2&$-1$&0 \\ \hline
$R_3^{(m,m)t}$&3&$\chi_3^{ii}$&$\chi_3^{ij}$&0&
$(-1)^t\omega_n^{ml}$ \\ \hline
$R_6^{(m_1,m_2)}$&6&$\chi_6^{ii}$&$\chi_6^{ij}$&0&0 \\ \hline
\end{tabular}
$(0<m< n,\,t \in {\bf Z}_2,\,
2 m_2-m_1 \geq 0,\,
m_2-2m_1 > -n,\,
m_1-m_2 > 0)$
\begin{eqnarray*}
\chi_3^{ij}&=&\omega_n^{m(i+j)}+\omega_n^{-mi}+\omega_n^{-mj} \\
\chi_6^{ij}&=&\omega_n^{m_1 i +m_2 j}
+\omega_n^{-m_1 (i+j)+m_2 i}
+\omega_n^{m_1 j -m_2 (i+j)} \\
&&+\omega_n^{m_1 j +m_2 i}
+\omega_n^{-m_1 (i+j)+m_2 j}
+\omega_n^{m_1 i -m_2 (i+j)}
\end{eqnarray*}
\begin{eqnarray*}
&&C_1 = \{A_{0,0}\} \\
&&C_2^i = \{A_{i,i}, A_{-2i,i}, A_{i,-2i}\} \quad (0 < i < n) \\
&&C_3^{ij} = \{A_{i,j}, A_{-i-j,i}, A_{j,-i-j},
A_{j,i}, A_{-i-j,j}, A_{i,-i-j}\} \quad
(i+2j > n,\, 2i+j < 2n,\, i-j > 0) \\
&&C_4 = \{C_{i,j}, E_{i,j}:0 \leq i,j < n\} \\
&&C_5^l = \{B_{i,l-i}, D_{i,n/2-l}, F_{n/2-l,j}:0
\leq i,j < n\} \quad (0 \leq l < n)
\end{eqnarray*}
\end{center}

\vskip 0.5cm

\begin{center}
\begin{tabular}{|c|c|c|c|c|c|}
\multicolumn{6}{c}
{Character table for
$\Delta(6n^2) \quad (n \in 3{\bf Z})$}\\
\hline
&$C_1^l$&$C_2^i$&$C_3^{ij}$&$C_4^l$&$C_5^l$ \\ \hline
$|C_a|$&1&3&6&$2n^2/3$&$3n$ \\ \hline
$\#$classes&3&$n-3$&$(n^2-3n+6)/6$&3&$n$ \\ \hline
$R_1^t$&1&1&1&1&$(-1)^t$ \\ \hline
$R_2^0$&2&2&2&$-1$&0 \\ \hline
$R_2^k$&2&2&$\omega_3^{i-j}+\omega_3^{j-i}$&
$\omega_3^{k+l}+\omega_3^{-k-l}$&0 \\ \hline
$R_3^{(m,m)t}$&$3\omega_3^{2lm}$&$\chi_3^{ii}$&
$\chi_3^{ij}$&0&$(-1)^t\omega_n^{ml}$ \\ \hline
$R_6^{(m_1,m_2)}$&$6\omega_3^{l(m_1+m_2)}$&$\chi_6^{ii}$&
$\chi_6^{ij}$&0&0 \\ \hline
\end{tabular}
$(0 \leq k \leq 2,\,0<m< n,\,t \in {\bf Z}_2,\,
2 m_2-m_1 \geq 0,\,
m_2-2m_1 > -n,\,
m_1-m_2 > 0)$
\begin{eqnarray*}
&&C_1^l = \{A_{ln/3,ln/3}\} \quad (l=0,1,2) \\
&&C_2^i = \{A_{i,i}, A_{-2i,i}, A_{i,-2i}\} \quad (0<i<n,\,
i \neq n/3,2n/3) \\
&&C_3^{ij} = \{A_{i,j}, A_{-i-j,i}, A_{j,-i-j},
A_{j,i}, A_{-i-j,j}, A_{i,-i-j}\} \quad
(i+2j > n,\, 2i+j < 2n,\, i-j > 0) \\
&&C_4^l = \{C_{i,i+3{\bf Z}+l},E_{i+3{\bf Z}+l,i}:0 \leq i < n \}
\quad (l=0,1,2) \\
&&C_5^l = \{B_{i,l-i}, D_{i,n/2-l}, F_{n/2-l,j}:0
\leq i,j < n\} \quad (0 \leq l < n)\end{eqnarray*}
\end{center}

\end{document}